\documentclass[epj,final]{svjour}
\usepackage{amsmath}
\usepackage{amssymb}
\usepackage{graphics}
\usepackage[unicode=true,pdfusetitle,
 bookmarks=true,bookmarksnumbered=false,bookmarksopen=false,
 breaklinks=false,pdfborder={0 0 1},backref=false,colorlinks=false]
 {hyperref}

\begin{document}

\title{Long-term effect of inter-mode transitions in quantum Markovian process}

\author{Sheng-Wen Li\inst{1} \and Li-Ping Yang\inst{2} \and C. P. Sun\inst{1}}

\institute{Beijing Computational Science Research Center, Beijing 100084, China
	\and Institute of Theoretical Physics, Chinese Academy of Sciences, 
		Beijing 100190, China}
\abstract{
We study the Markovian process of a multi-mode open system connecting
with a non-equilibrium environment, which consists of several heat
baths with different temperatures. As an illustration, we study the
steady state of three linearly coupled harmonic oscillators in long
time evolution, two of which contact with two independent bosonic
heat baths with different temperatures respectively. We show that
the inter-mode transitions mediated by the environment is responsible
for the long time behavior of the dynamics evolution, which is usually
considered to take effect only in short time dynamics of the system
immersed in a equilibrium heat bath with a single  temperature. These
inter-mode transitions are essential to the non-equilibrium flux between
subsystems, thus they cannot be neglected.
}

\maketitle

\section{Introduction}

When a small system contacts with a simple environment, i.e., a heat
bath in canonical equilibrium with a  temperature $T$, it would approach
its canonical state with the same temperature $T$. This dynamics
process is called \emph{thermalization} \cite{breuer_theory_2002,rigol_thermalization_2008,linden_quantum_2009,liao_single-particle_2010,liao_quantum_2011}.
However, non-equilibrium systems are more general in nature, and  exhibit
more rich physics. A typical example is a composite system connecting
with more than one heat baths with different temperatures. For a long-term
evolution, the open composite system would not approach its canonical
thermal state, but still it would be stabilized to a certain steady
state. We call this process \emph{non-thermal stabilization}.

Such composite system coupling with multiple independent heat baths
appear in many artificial systems, like the superconducting circuit
and quantum dots, and also natural systems, like the excitons in photon-synthesis
system \cite{caruso_highly_2009,yang_dimerization-assisted_2010,liao_coherent_2010}.
In these composite systems, the interaction between the subsystems
is always on, and that may affect the response of the system to the
environment.

A rigorous treatment of the interacting composite system should be
based on the normal modes of the system. In an open system, both the
equilibrium and non-equilibrium case as we mentioned above, the energy
exchange with environment would mediate the transitions between these
normal modes of the total system, which we call the \emph{inter-mode
transition}. It was usually believed that this inter-mode transition
only takes effect on the dynamics of transient evolution within the
time scale determined by the time-energy uncertainty \cite{breuer_theory_2002,jing_breakdown_2009,li_effect_2009,ai_quantum_2010,li_collective_2012},
and averagely it has no effect to the steady state behavior after
a long time evolution. This is also known as secular approximation
or rotating-wave approximation (RWA).

However, in this paper, we find that indeed such inter-mode transitions
 have long-term effect in non-equilibrium system even for Markovian
process. As an example, we study the steady state of three linearly
coupled harmonic oscillators (HOs), two of which contact with two
independent bosonic heat baths with different temperatures respectively.
We find that if the inter-mode transition were ignored, there would
be some counter-intuitive results in the long time steady state. We
show that the inter-mode transitions are essential to the non-equilibrium
flux inside the composite system. As a comparison, we also show that
such effect does not appear in equilibrium environments. We emphasize
that the omission of these inter-mode transitions is consistent with
conventional equilibrium reservoirs as studied in previous works \cite{breuer_theory_2002,jing_breakdown_2009,li_effect_2009,ai_quantum_2010,li_collective_2012}. 

The paper is arranged as follows. In Sec.\,II, we setup the model
of the coupled system and give a master equation. In Sec.\,III, we
give the stabilization result and make some analytical discussion
by eliminating the degree of freedom of the mediating data bus. We
show that the omission of the inter-mode transition is consistent
with the equilibrium reservoirs, and give a physical explanation.
In Sec.\,IV, We propose a possible implementation. The calculation
is assisted by some properties of the characteristic description of
Wigner function and Fokker-Planck equation. We leave these tricks
in the appendices. Finally, summary is drawn in Sec.\,V.

\section{Model setup}

To study the long-term dynamics of a composite system coupled to a
complicated environment, we use the coupled HOs system as an illustration.
The system we study here is illustrated in Fig.\,1. Two HOs with
frequencies $\omega_{L,R}$ contact with two independent heat baths
with different temperatures. In experiments, microscopic devices with
mutual interactions are separated from each other for only several
micrometers, thus it is unclear to discuss their local temperatures.
Here we introduce a third HO as a data bus to mediate their coupling, which makes
it possible to separate the two HOs for a certain distance and we
can discuss their local temperatures clearly. Effectively, we suppose
the mediating HO does not contact with any environment. 

The three oscillators system can be described by a quadratic coupled
Hamiltonian $H_{S}=H_{0}+V$, where
\begin{align}
H_{0} & =\omega_{L}\hat{a}_{L}^{\dagger}\hat{a}_{L}+\omega_{R}\hat{a}_{R}^{\dagger}\hat{a}_{R}+\omega_{\mathrm{m}}\hat{b}^{\dagger}\hat{b},\label{eq:H_S}\\
V & =g_{L}(\hat{a}_{L}^{\dagger}\hat{b}+\hat{a}_{L}\hat{b}^{\dagger})+g_{R}(\hat{a}_{R}^{\dagger}\hat{b}+\hat{a}_{R}\hat{b}^{\dagger}),\nonumber 
\end{align}
and $H_{0}$ describes the free Hamiltonian with local modes respectively
defined by annihilation operators $\hat{a}_{L},\,\hat{a}_{R}$ and
$\hat{b}$; $V$ describes the coupling among the local modes.

\begin{figure}
	\begin{center}
	\resizebox{0.4\textwidth}{!}{
		\includegraphics{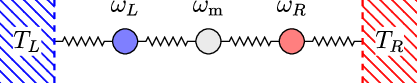}
		}
	\end{center}
\caption{(Color online) Demonstration of the coupled oscillators system. Two
remotely located HOs are indirectly coupled by another mediating one.
The HOs at the two ends contact with independent heat baths with different
temperatures $T_{L/R}$.}
\end{figure}

We assume the two oscillators locate remotely at different places,
thus they may suffer from independent baths. We also assume each bath
stays at a canonical thermal state with a  temperature $T_{L/R}$.
The whole system can be described by the total Hamiltonian ${\cal H}=H_{S}+H_{B}+V_{SB}$,
where 
\begin{align}
H_{B} & =\sum_{\mathbf{k}_{L}}\omega_{\mathbf{k}_{L}}\hat{c}_{\mathbf{k}_{L}}^{\dagger}\hat{c}_{\mathbf{k}_{L}}+\sum_{\mathbf{k}_{R}}\omega_{\mathbf{k}_{R}}\hat{c}_{\mathbf{k}_{R}}^{\dagger}\hat{c}_{\mathbf{k}_{R}},\nonumber \\
V_{SB} & =\sum_{\sigma=L,R}\hat{a}_{\sigma}^{\dagger}\Gamma_{\sigma}+\hat{a}_{\sigma}\Gamma_{\sigma}^{\dagger},
\end{align}
 and $\Gamma_{\sigma}=\sum_{\mathbf{k}_{\sigma}}g_{\mathbf{k}_{\sigma}}\,\hat{c}_{\mathbf{k}_{\sigma}}$.
$H_{B}$ is the free Hamiltonian of the two boson heat baths, each
of which is modeled as a collection of boson modes, described by the
boson annihilation operators $\hat{c}_{\mathbf{k}_{\sigma}}$. $V_{SB}$
represents the linear coupling between the system and the environment.

We need to derive a master equation to study the dynamics of the open
system. Actually for the coupled HO system, a correct treatment of
the master equation should be based on the normal modes of $H_{S}$,
but not the local modes $\hat{a}_{L/R}$ and $\hat{b}$. Otherwise,
it may give rise to some counter-intuitive results. Thus, we diagonalize
the Hamiltonian $H_{S}$ as,
\begin{align}
H_{S} & =(\hat{a}_{L}^{\dagger},\,\hat{b}^{\dagger},\,\hat{a}_{R}^{\dagger})\left[\begin{array}{ccc}
\omega_{L} & g_{L}\\
g_{L} & \omega_{\mathrm{m}} & g_{R}\\
 & g_{R} & \omega_{R}
\end{array}\right]\left(\begin{array}{c}
\hat{a}_{L}\\
\hat{b}\\
\hat{a}_{R}
\end{array}\right)\nonumber \\
 & \equiv\mathbf{a}{}^{\dagger}\cdot\Omega\cdot\mathbf{a}=\sum_{i=1}^{3}\varepsilon_{i}\hat{A}_{i}^{\dagger}\hat{A}_{i},\label{eq:coupled-H}
\end{align}
 where \textbf{$\mathbf{a}=(\hat{a}_{L},\,\hat{b},\,\hat{a}_{R})^{T}$,}
and we also denote it as $(\hat{a}_{1},\,\hat{a}_{2},\,\hat{a}_{3})^{T}$
hereafter (with redefined indices $1,2,3$ for $\hat{a}_{L},\,\hat{b},\,\hat{a}_{R}$
respectively). $\mathbf{A}=U\cdot\mathbf{a}=(\hat{A}_{1},\,\hat{A}_{2},\,\hat{A}_{3})^{T}$
for $A_{i}$'s being the normal modes. $U\cdot\Omega\cdot U^{\dagger}=\mathrm{diag}\left\{ \varepsilon_{1},\varepsilon_{2},\varepsilon_{3}\right\} $
gives the eigen frequencies of the normal modes. Although the normal
modes are decoupled from each other in the isolated $H_{S}$, we can
see below that the environment could mediately induce some effective
couplings between these normal modes.

With the above notations, in Appendix A we derive a master equation
to describe the long-term dynamics of the open system via Born-Markovian
approximation \cite{breuer_theory_2002}. In Schr\"odinger's picture,
it reads as 
\begin{align}
\partial_{t}\rho= & i[\rho,\,\sum\varepsilon_{i}\hat{A}_{i}^{\dagger}\hat{A}_{i}]\nonumber \\
 & +\sum_{ij}\,\frac{\Lambda_{ij}^{-}}{2}\big(2\hat{A}_{i}\rho\hat{A}_{j}^{\dagger}-\{\hat{A}_{j}^{\dagger}\hat{A}_{i},\rho\}_{+}\big)\label{eq:ME-noRWA}\\
 & +\frac{\Lambda_{ij}^{+}}{2}\big(2\hat{A}_{i}^{\dagger}\rho\hat{A}_{j}-\{\hat{A}_{j}\hat{A}_{i}^{\dagger},\rho\}_{+}\big),\nonumber 
\end{align}
where 
\begin{align}
\Lambda_{ij}^{+}= & \frac{\gamma_{L}}{2}U_{i1}U_{j1}^{*}[N_{L}(\varepsilon_{i})+N_{L}(\varepsilon_{j})]\nonumber \\
 & +\frac{\gamma_{R}}{2}U_{i3}U_{j3}^{*}[N_{R}(\varepsilon_{i})+N_{R}(\varepsilon_{j})],\\
\Lambda_{ji}^{-}= & \frac{\gamma_{L}}{2}U_{i1}U_{j1}^{*}[N_{L}(\varepsilon_{i})+N_{L}(\varepsilon_{j})+2]\nonumber \\
 & +\frac{\gamma_{R}}{2}U_{i3}U_{j3}^{*}[N_{R}(\varepsilon_{i})+N_{R}(\varepsilon_{j})+2].\nonumber 
\end{align}
Here, $\gamma_{\sigma}(\varepsilon_{i})=2\pi J_{\sigma}(\varepsilon_{i})$
characterizes the coupling strength with each bath, and $J_{\sigma}(\omega)=\sum_{\mathbf{k_{\sigma}}}\left|g_{\mathbf{k}_{\sigma}}\right|^{2}\delta(\omega-\omega_{\mathbf{k}_{\sigma}})$
is the coupling distribution. For the usual case, we can assume that
$\gamma_{\sigma}(\omega)\simeq\gamma_{\sigma}$ does not depend too
much on $\omega$ and can be treated as constant. $N_{\sigma}(\omega)=[\exp(\omega/kT_{\sigma})-1]^{-1}$
is the Planck distribution for $\sigma=L,R$.

It is observed from the above master equation that the environment
indeed induces an effective coupling between two normal modes $A_{i}$
and $A_{j}$. $\Lambda_{ij}^{\pm}$ measure the transitions of the
normal modes. This environment-mediating effect can be understood
in the following way. The coupled system exchanges energy with environment
through the interaction $V_{SB}$. Immediately after the normal mode
$A_{i}$ of the system emits an energy quanta $\varepsilon_{i}$ to
the environment, another process may happen in succession that the
normal mode $A_{j}$ absorbs back $\varepsilon_{j}$ from the environment.
Also, there is another possibility for the reversed process. Thus,
different normal modes $A_{i}$'s of $H_{S}$ are coupled with the
mediation of the environment. In fact, this effect of environment
mediated coupling can be also found for two modes coupled to a common
heat bath \cite{mccutcheon_long-lived_2009,ma_entanglement_2012}.

The transition terms with $i\neq j$ describes the effective coupling
between different normal modes. In the interaction picture, these
terms would contribute an oscillating factor $\exp[\pm i\delta\epsilon_{ij}\, t]$
resulted from the energy difference of the modes $\hat{A}_{i}$ and
$\hat{A}_{j}$. This transition effect would be ignored if we apply
RWA by dropping these terms. However, as can be seen in the following,
such ignorance would give rise to counter-intuitive results for non-equilibrium
system.

\section{Long-term stabilization dynamics}

\begin{figure*}
	\begin{center}
	\resizebox{0.95\textwidth}{!}{
		\includegraphics{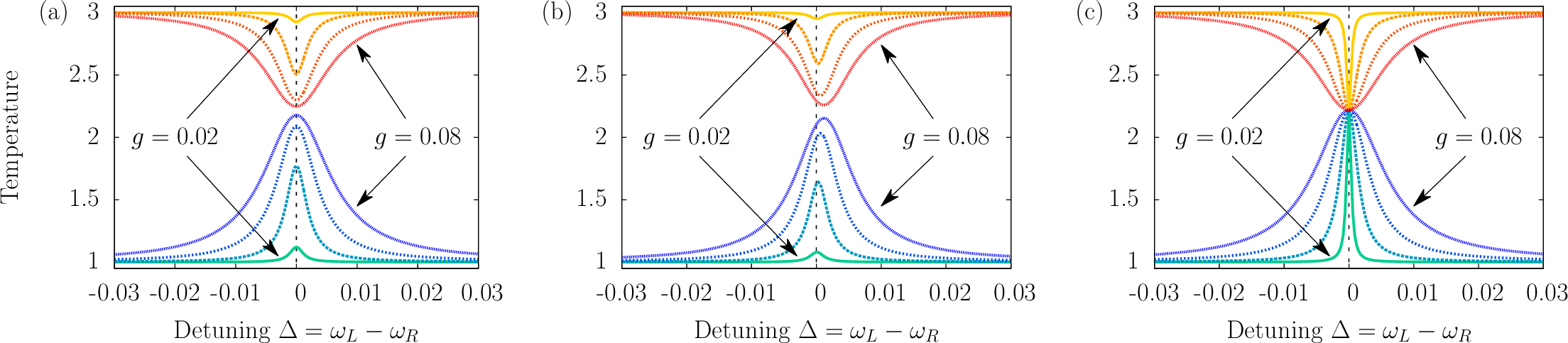}
		} 
	\end{center}
\caption{(Color online) The effective temperatures $T_{L}^{\mathrm{eff}}$
(lower blue ones) and $T_{R}^{\mathrm{eff}}$ (upper red ones) of
the oscillators at the two ends calculated from the result without
(a) (b) and with (c) RWA. We set $\overline{\omega}=(\omega_{L}+\omega_{R})/2=1$
as the unit, and $T_{L}=1,\, T_{R}=3,\,\gamma_{L}=0.002,\,\gamma_{R}=0.003,\,\omega_{\mathrm{m}}=2$.
We set $g_{L}=g_{R}=g$ in (a) (c), and $g_{L}=g,\, g_{R}=0.8g$ in
(b). We plot four groups of curves according to $g=0.02,\,0.04,\,0.06,\,0.08$,
distributed from outside to inside in these figures. At the small
regime around the degeneracy point, RWA cannot give us a good enough
result, especially when $g$ is weak. The extremum points are shifted
aside when $g_{L}\neq g_{R}$.}
\end{figure*}

Comparing with the long time Markovian thermalization process in a
heat bath with a single temperature, the present environment with
two temperatures $T_{L/R}$ cannot stabilize the whole system into
a canonical thermal state. In this section, we first calculate the
steady state of the open quantum system by straightforwardly solving
the above master equation Eq.\,(\ref{eq:ME-noRWA}). Then we consider
the mediating HO as a quantum data bus in the large detuning limit.
The adiabatic elimination of this oscillator can formally induce a
direct coupling between the left and right HOs. In this case, the
analytical results about the stabilization can be obtained explicitly.

\subsection{Steady state in long time limit}

We now consider the indirect coupling case with a mediating data bus.
The master equation without RWA can be solved with the help of the
characteristic function of Wigner representation, which is defined
as (see Appendix B),
\begin{align}
\chi(\vec{\mu}) & \equiv\mathbf{Tr}\big[\rho\cdot\exp(\mathbf{A}^{\dagger}\cdot\vec{\mu}-\vec{\mu}^{\dagger}\cdot\mathbf{A})\big]\\
 & =\mathbf{Tr}\big[\rho\cdot\exp(\mathbf{a}^{\dagger}\cdot\vec{\kappa}-\vec{\kappa}^{\dagger}\cdot\mathbf{a})\big].\nonumber 
\end{align}
Here, $\vec{\mu}=(\mu_{1},\mu_{2},\mu_{3})^{T}$ and $\vec{\kappa}=(\kappa_{1},\kappa_{2},\kappa_{3})^{T}$
are complex vectors with respect to the normal and local modes, and
$\vec{\mu}=U\cdot\vec{\kappa}$. The corresponding Wigner function
with three modes is defined as the Fourier transform of $\chi(\vec{\kappa})$,
\[
W(\vec{\alpha},\vec{\alpha}^{*})=\frac{1}{(\pi^{2})^{3}}\int d^{2}\vec{\kappa}\, e^{-\vec{\alpha}^{\dagger}\cdot\vec{\kappa}+\vec{\kappa}^{\dagger}\cdot\vec{\alpha}}\chi(\vec{\kappa}).
\]

With this definition, we obtain the equation of $\chi(\vec{\mu})$
as
\begin{gather}
\partial_{t}\chi+\mathbf{z}\cdot\mathbf{T}\cdot\frac{\partial}{\partial\mathbf{z}^{T}}\,\chi=\mathbf{z}\cdot\mathbf{D}\cdot\mathbf{z}^{T}\,\chi,\label{eq:FP-equation}
\end{gather}
 where $\mathbf{z}=(\mu_{1},\mu_{2},\mu_{3},\,\mu_{1}^{*},\mu_{2}^{*},\mu_{3}^{*})$,
and
\[
\mathbf{T}=\left[\begin{array}{cc}
T^{-} & \mathbf{0}\\
\mathbf{0} & T^{+}
\end{array}\right],\quad\mathbf{D}=\left[\begin{array}{cc}
\mathbf{0} & P\\
P^{T} & \mathbf{0}
\end{array}\right].
\]
 \textbf{$\mathbf{T}$} and $\mathbf{D}$ are $6\times6$ matrices
with $3\times3$ blocks $T^{\pm}$ and $P$ defined by 
\begin{align}
P_{ij} & =-\frac{1}{4}(\Lambda_{ij}^{-}+\Lambda_{ji}^{+}),\nonumber \\
T_{ij}^{-} & =\frac{1}{2}(\Lambda_{ij}^{-}-\Lambda_{ji}^{+})-i\varepsilon_{i}\delta_{ij},\\
T_{ij}^{+} & =\frac{1}{2}(\Lambda_{ji}^{-}-\Lambda_{ij}^{+})+i\varepsilon_{i}\delta_{ij}.\nonumber 
\end{align}

The equation (\ref{eq:FP-equation}) is the Fourier transformation
of the Fokker-Planck equation about the Wigner function \cite{walls_quantum_2008}.
Formally we give the analytical solution for the steady state in Appendix
C. Its expression is given as,
\begin{gather}
\chi(\vec{\mu})=\chi(U\cdot\vec{\kappa})=\exp\left[\mathbf{z}V^{-1}\cdot\mathbf{D}'\cdot(\mathbf{z}V^{-1})^{T}\right],\label{eq:steady}
\end{gather}
where $V$ diagonalizes the matrix $\mathbf{T}$, i.e., $V\cdot\mathbf{T}\cdot V^{-1}=\mathrm{diag}\{\mathbf{d}_{1},\cdots,\mathbf{d}_{6}\}$,
and $\mathbf{D}'_{ij}=[V\mathbf{D}V^{T}]_{ij}/(\mathbf{d}_{i}+\mathbf{d}_{j}).$

All the steady state properties of the composite system can be obtained
from this formal solution Eq.\,(\ref{eq:steady}). Specially, we
are interested in the steady state of the HOs at the two ends. We
can obtain $\chi_{\sigma}(\kappa_{\sigma})$ for each local oscillator
just by setting $\kappa_{i}=0$ for all $i\neq\sigma$. Notice that
$V^{-1}$ and $\mathbf{D}'$ in the exponent of Eq.\,(\ref{eq:steady})
are block diagonal and anti-diagonal respectively, thus it can be
verified that $\chi_{\sigma}(\kappa_{\sigma})$ is always of the following
Gaussian form, 
\begin{equation}
\chi(\kappa_{\sigma},\kappa_{\sigma}^{*})=\exp\big[-(N_{\sigma}^{\mathrm{eff}}+\frac{1}{2})\left|\kappa_{\sigma}\right|^{2}\big],\label{eq:gauss}
\end{equation}
where $N_{\sigma}^{\mathrm{eff}}$ is a positive constant. In Appendix
B, we show that if $\chi(\kappa_{\sigma},\kappa_{\sigma}^{*})$ has
the Gaussian form like Eq.\,(\ref{eq:gauss}), the state of the oscillator
is a canonical state, and there is no squeezing. Since each HO can finally reach a canonical
steady state, we can treat it as an equivalent thermal state and define
an effective temperature from its average occupation $N_{\sigma}^{\mathrm{eff}}=\langle\hat{a}_{\sigma}^{\dagger}\hat{a}_{\sigma}\rangle$,
\begin{equation}
T_{\sigma}^{\mathrm{eff}}=\omega_{\sigma}/\ln(1+\frac{1}{N_{\sigma}^{\mathrm{eff}}}).
\end{equation}

In Fig.\,2(a, b) we show the effective temperatures of the oscillators
at the two ends calculated from Eq.\,(\ref{eq:steady}), changing
with the detuning $\Delta=\omega_{L}-\omega_{R}$ and the coupling
strengths $g_{L/R}$. Here we set $\overline{\omega}=(\omega_{L}+\omega_{R})/2\equiv1$
as the energy scale and $\omega_{\mathrm{m}}=2$. 

When the detuning $\Delta$ becomes large or when their coupling strength
$g_{L/R}$ becomes small, the two oscillators tend to be thermalized
with their own heat bath respectively. The effective temperatures
get to the closest point around the resonance regime $\Delta\simeq0$.
This observation means that they are affected by the reservoir at
the opposite side and heat transfer happens greatly. When $g_{L}=g_{R}$,
the extremum points locate exactly at $\omega_{L}=\omega_{R}$, while
they shift aside when $g_{L}\neq g_{R}$. When the interaction becomes
strong, the effective temperatures of the two oscillators tend to
get closer and closer, away from that of each heat bath.

As comparison, we also show some counter-intuitive observation resulting
from the improper omission of the transition terms like $2\hat{A}_{i}\rho\hat{A}_{j}^{\dagger}-\{\hat{A}_{j}^{\dagger}\hat{A}_{i},\rho\}$
with $i\neq j$ {[}see Fig.\,2(c){]}. In the large detuning area,
this approximation shows well consistence with previous result in
Fig.\,2(a). But the effective temperatures always equals at the degeneracy
point even when the coupling strength $g$ is quite weak, i.e., when
the oscillators tend to be decoupled from each other. A similar problem
was also studied in Ref.\,\cite{liao_quantum_2011}, where they considered
two interacting two-level systems respectively contacting two independent
heat baths with different temperatures, and they obtained a result
similar to ours shown in Fig.\,2(c), which is valid only when the
coupling strength is quite large.

With the above comparison, we look back at the master equation carefully,
the transition terms like $e^{-i\delta\epsilon_{ij}\, t}\times\big(2\hat{A}_{i}\rho\hat{A}_{j}^{\dagger}-\{\hat{A}_{j}^{\dagger}\hat{A}_{i},\rho\}_{+}\big)$
with $i\neq j$ contribute to the energy exchange of different modes
$\hat{A}_{i}$ and $\hat{A}_{j}$, which can be characterized by $\langle\hat{A}_{i}^{\dagger}\hat{A}_{j}\rangle$.
The oscillating factor at the front describes the phase of this transition.
The transition rate $\delta\epsilon_{ij}=\varepsilon_{i}-\varepsilon_{j}$
is determined by the detuning and coupling strength. When $\Delta$
and $g_{L,R}$ are quite small, the omission of these terms seems
doubtable.

Intuitively, the only reason why these inter-mode transition terms
cannot be dropped is that they rotate too slowly. However, remember
that we only focus on the steady behavior $t\rightarrow\infty$. In
this case, even a quite slowly rotating term should be averaged to
zero. Indeed, in the following we would see that in equilibrium reservoirs,
ignorance of such transitions does give the correct result even when
the transition rate $\delta\epsilon_{ij}$ is small, and the real
reason lies in the non-equilibrium environment.

\subsection{Effective coupling in adiabatic limit}

When the detuning of $\omega_{\mathrm{m}}$ to $\omega_{L/R}$ is
large, we can eliminate the mediating degree of freedom adiabatically
to simplify our analysis. We apply Fr\"ohlich-Nakajima transformation
\cite{frohlich_theory_1950,nakajima_perturbation_1955,xiang_hybrid_2012},
and obtain the following simplified Hamiltonian, which describes a
system of two directly coupled oscillators,
\begin{equation}
H_{S}=\omega_{L}'\,\hat{a}_{L}^{\dagger}\hat{a}_{L}+\omega_{R}'\,\hat{a}_{R}^{\dagger}\hat{a}_{R}+g(\hat{a}_{L}^{\dagger}\hat{a}_{R}+\hat{a}_{L}\hat{a}_{R}^{\dagger}),
\end{equation}
where
\begin{align}
\omega_{L,R}' & =\omega_{L,R}+\frac{g_{L,R}^{2}}{\omega_{L,R}-\omega_{\mathrm{m}}},\label{eq:renormal}\\
g & =\frac{1}{2}\big(\frac{g_{L}g_{R}}{\omega_{L}-\omega_{\mathrm{m}}}+\frac{g_{L}g_{R}}{\omega_{R}-\omega_{\mathrm{m}}}\big).\nonumber 
\end{align}
 The coupling strengths $g_{L}$ and $g_{R}$ contribute a correction
to the renormalized frequencies $\omega_{L,R}'$. 

For this simplified Hamiltonian, we can write down the analytical
expression of eigen frequencies $\varepsilon_{i}$ and the transformation
$U$ for the normal modes $\hat{A}_{\pm}$. Denoting $\omega_{L}'=\overline{\omega}-\Delta/2,\,\omega_{R}'=\overline{\omega}+\Delta/2$,
we have
\begin{align}
\varepsilon_{\pm} & =\overline{\omega}\pm\tilde{\Delta}_{g},\qquad\tilde{\Delta}_{g}=(\frac{\Delta^{2}}{4}+g^{2})^{\frac{1}{2}},\nonumber \\
U & =\left[\begin{array}{cc}
\alpha & \beta\\
\beta & -\alpha
\end{array}\right],\quad\frac{2\alpha\beta}{\alpha^{2}-\beta^{2}}=\frac{2g}{\Delta}.\label{eq:Delta_g}
\end{align}
It follows from Eq.\,(\ref{eq:Delta_g}) that the energy difference
$\delta\epsilon_{ij}=2\tilde{\Delta}_{g}$ depends on the detuning
$\Delta$ and coupling strength $g$. When $\Delta$ and $g$ are
small, the factors $\exp[\pm2i\tilde{\Delta}_{g}t]$ of the transition
terms between the two normal modes oscillate quite slowly.

We carry out the similar calculation for this simplified two oscillators
system as previously, which gives an explicit expression for the steady
state of each oscillator, described by the characteristic function
$\chi_{\sigma}(\kappa_{\sigma})$, 
\begin{align}
\chi_{\sigma}(\kappa_{\sigma})= & \exp\big[-(N_{\sigma}^{\mathrm{eff}}+\frac{1}{2})\left|\kappa_{\sigma}\right|^{2}\big],\label{eq:steady-each}\\
N_{\sigma}^{\mathrm{eff}}= & \big[\mathsf{A}_{\sigma}N_{L}(\varepsilon_{-})+\mathsf{B}_{\sigma}N_{L}(\varepsilon_{+})\nonumber \\
 & +\mathsf{C}_{\sigma}N_{R}(\varepsilon_{-})+\mathsf{D}_{\sigma}N_{R}(\varepsilon_{+})\big]/\Phi.\nonumber 
\end{align}
Here $N_{\sigma}^{\mathrm{eff}}$ is the occupation number of the
effective thermal distribution ($\sigma=L,R$) determined by the linear
combination of $N_{L,R}(\varepsilon_{\pm})$, and the coefficients
are,
\[
\Phi=\gamma_{L}\gamma_{R}(\gamma_{L}+\gamma_{R})^{2}+16\tilde{\Delta}_{g}^{2}(\alpha^{2}\gamma_{L}+\beta^{2}\gamma_{R})(\beta^{2}\gamma_{L}+\alpha^{2}\gamma_{R}),
\]
\begin{align*}
\mathsf{A}_{L} & =\alpha^{2}\big[\gamma_{L}\gamma_{R}(\gamma_{L}+\gamma_{R})^{2}+16\alpha^{2}\tilde{\Delta}_{g}^{2}(\beta^{2}\gamma_{L}+\alpha^{2}\gamma_{R})\gamma_{L}\big],\\
\mathsf{B}_{L} & =\beta^{2}\big[\gamma_{L}\gamma_{R}(\gamma_{L}+\gamma_{R})^{2}+16\beta^{2}\tilde{\Delta}_{g}^{2}(\alpha^{2}\gamma_{L}+\beta^{2}\gamma_{R})\gamma_{L}\big],\\
\mathsf{C}_{L} & =16\alpha^{2}\beta^{2}\tilde{\Delta}_{g}^{2}(\beta^{2}\gamma_{L}+\alpha^{2}\gamma_{R})\gamma_{R},\\
\mathsf{D}_{L} & =16\alpha^{2}\beta^{2}\tilde{\Delta}_{g}^{2}(\alpha^{2}\gamma_{L}+\beta^{2}\gamma_{R})\gamma_{R},
\end{align*}
and
\begin{align*}
\mathsf{A}_{R} & =16\alpha^{2}\beta^{2}\tilde{\Delta}_{g}^{2}(\beta^{2}\gamma_{L}+\alpha^{2}\gamma_{R})\gamma_{L},\\
\mathsf{B}_{R} & =16\alpha^{2}\beta^{2}\tilde{\Delta}_{g}^{2}(\alpha^{2}\gamma_{L}+\beta^{2}\gamma_{R})\gamma_{L},\\
\mathsf{C}_{R} & =\beta^{2}\big[\gamma_{L}\gamma_{R}(\gamma_{L}+\gamma_{R})^{2}+16\beta^{2}\tilde{\Delta}_{g}^{2}(\alpha^{2}\gamma_{L}+\beta^{2}\gamma_{R})\gamma_{R}\big],\\
\mathsf{D}_{R} & =\alpha^{2}\big[\gamma_{L}\gamma_{R}(\gamma_{L}+\gamma_{R})^{2}+16\alpha^{2}\tilde{\Delta}_{g}^{2}(\beta^{2}\gamma_{L}+\alpha^{2}\gamma_{R})\gamma_{R}\big].
\end{align*}

From Eq.\,(\ref{eq:steady-each}) we see that each oscillator achieves
a canonical state. Especially, at the degeneracy point $\omega_{L}'=\omega_{R}'$,
we have $\alpha^{2}=\beta^{2}=1/2$, and the difference of the populations
is, 
\begin{multline}
N_{L}^{\mathrm{eff}}-N_{R}^{\mathrm{eff}}\\
=\cfrac{\big[N_{L}(\varepsilon_{+})-N_{R}(\varepsilon_{+})\big]+\big[N_{L}(\varepsilon_{-})-N_{R}(\varepsilon_{-})\big]}{2(1+4g^{2}/\gamma_{L}\gamma_{R})}.\label{eq:N_L-N_R}
\end{multline}
 The above equation (\ref{eq:N_L-N_R}) explicitly shows that the
effective temperatures of the two oscillators are not equal at the
degeneracy point when $T_{L}\neq T_{R}$.

In the equilibrium case, we have $T_{L}=T_{R}=\overline{T}$ and $N_{L}(\varepsilon)=N_{R}(\varepsilon)\equiv\overline{N}(\varepsilon)$.
The above result Eq.\,(\ref{eq:steady-each}), which is obtained
without RWA, still holds. And we can explicitly obtain $N_{\sigma}^{\mathrm{eff}}$
as
\begin{align}
N_{L}^{\mathrm{eff}} & =\alpha^{2}\overline{N}(\varepsilon_{-})+\beta^{2}\overline{N}(\varepsilon_{+}),\nonumber \\
N_{R}^{\mathrm{eff}} & =\beta^{2}\overline{N}(\varepsilon_{-})+\alpha^{2}\overline{N}(\varepsilon_{+}).\label{eq:simple}
\end{align}

However, a simple calculation by omitting the inter-mode transitions
also gives exactly the same analytical result as Eq.\,(\ref{eq:simple}),
even when the transition rate $\tilde{\Delta}_{g}=[\Delta^{2}/4+g^{2}]^{1/2}$
is small. Both calculations, with and without RWA, give the steady
state of the two oscillators, i.e.,
\begin{equation}
\rho_{s}=\frac{1}{{\cal Z}}\exp[-\frac{1}{k\overline{T}}(\varepsilon_{-}\hat{A}_{-}^{\dagger}\hat{A}_{-}+\varepsilon_{+}\hat{A}_{+}^{\dagger}\hat{A}_{+})],
\end{equation}
 no matter how slowly the transition coefficients rotate.

\subsection{Inter-mode transition and flux}

Here we give an physical explanation why the omission of the inter-mode
transitions is consistent with equilibrium system but not allowed
for non-equilibrium system. We still consider the model of three oscillators.
If we omit all the inter-mode transitions in Eq.\,(\ref{eq:ME-noRWA}),
we obtain the following master equation,
\begin{align}
\partial_{t}\rho= & i[\rho,\,\sum\varepsilon_{i}\hat{A}_{i}^{\dagger}\hat{A}_{i}]\nonumber \\
 & +\sum_{i}\,\frac{\Lambda_{ii}^{-}}{2}\big(2\hat{A}_{i}\rho\hat{A}_{i}^{\dagger}-\{\hat{A}_{i}^{\dagger}\hat{A}_{i},\rho\}_{+}\big)\\
 & +\frac{\Lambda_{ii}^{+}}{2}\big(2\hat{A}_{i}^{\dagger}\rho\hat{A}_{i}-\{\hat{A}_{i}\hat{A}_{i}^{\dagger},\rho\}_{+}\big).\nonumber 
\end{align}
 In this equation with RWA, all the normal modes $\hat{A}_{i}$ are
decoupled from each other. It can be verified that the steady state
of this equation is
\begin{align}
\rho_{s} & =\frac{1}{{\cal Z}}\exp[-\sum_{i}\tilde{\beta}_{i}\hat{A}_{i}^{\dagger}\hat{A}_{i}],\nonumber \\
\tilde{\beta}_{i} & =\ln[\Lambda_{ii}^{-}/\Lambda_{ii}^{+}].
\end{align}
Such steady solution has a property that for $i\neq j$, we have $\mathrm{Tr}[\rho_{s}\hat{A}_{i}^{\dagger}\hat{A}_{j}]=0$,
which is also consistent with the fact that there is no inter-mode
transition.

Recall that $\hat{A}_{i}=U_{in}\hat{a}_{n}$, generally we can write
down the transition amplitudes for the local modes as
\[
\langle\hat{a}_{m}^{\dagger}\hat{a}_{n}\rangle=\sum_{i}U_{mi}U_{ni}^{*}\langle\hat{A}_{i}^{\dagger}\hat{A}_{i}\rangle+\sum_{i\neq j}U_{mi}U_{nj}^{*}\langle\hat{A}_{i}^{\dagger}\hat{A}_{j}\rangle.
\]
 Since we can always choose a proper phase to guarantee that all $U_{mi}$'s
are real, if all the inter-mode transitions are omitted, i.e., $\langle\hat{A}_{i}^{\dagger}\hat{A}_{j}\rangle=0$
for $i\neq j$, immediately we obtain 
\begin{equation}
\langle\hat{a}_{m}^{\dagger}\hat{a}_{n}\rangle-\langle\hat{a}_{n}^{\dagger}\hat{a}_{m}\rangle=0.
\end{equation}

Indeed, $\langle\hat{a}_{m}^{\dagger}\hat{a}_{n}\rangle-\langle\hat{a}_{n}^{\dagger}\hat{a}_{m}\rangle$
is proportional to the energy or particle flux between the local sites.
For example, we consider the particle exchange of the mediating mode
$\hat{b}$ shown in Fig.\,1. By Heisenberg equation, we have
\begin{align}
\partial_{t}\langle\hat{b}^{\dagger}\hat{b}\rangle= & ig_{L}(\langle\hat{a}_{L}^{\dagger}\hat{b}\rangle-\langle\hat{a}_{L}\hat{b}^{\dagger}\rangle)\nonumber \\
 & \quad+ig_{R}(\langle\hat{a}_{R}^{\dagger}\hat{b}\rangle-\langle\hat{a}_{R}\hat{b}^{\dagger}\rangle).
\end{align}
From this equation, we can define the particle flux from $\hat{b}$
to the left/right site as $J_{\sigma}\equiv ig_{\sigma}(\langle\hat{a}_{\sigma}^{\dagger}\hat{b}\rangle-\langle\hat{a}_{\sigma}\hat{b}^{\dagger}\rangle)$,
where $\sigma=L,R$. Thus, the omission of the inter-mode transitions
would always give $J_{\sigma}=0$, which means that there is no net
flux between the local sites.

For equilibrium systems, there is no net flux between the subsystems,
thus the omission of these inter-mode transitions is consistent, even
when the transition rate is quite small. That is, as we mentioned
before, when we focus on the steady behavior $t\rightarrow\infty$,
even a quite slowly rotating term should be averaged to zero. However,
the existence of a steady flux is an essential element of non-equilibrium
systems. Therefore, we conclude that for the non-equilibrium case,
the inter-mode transitions would contribute to long-term effect even
in Markovian systems. This is different from the previous viewpoints
that the inter-mode transitions only have transient effect within
the time scale determined by the time-energy uncertainty, which usually
applies in conventional thermalization process \cite{breuer_theory_2002,jing_breakdown_2009,li_effect_2009,ai_quantum_2010,li_collective_2012}.

\section{Physical implementation}

In this section, we discuss a possible implementation scheme, which
is composed of two nano-mechanical oscillators (NAMR) connected via
a superconducting transmission line (TLR), in order to test the theoretical
results we have got here, as shown in Fig.\,3.

The electromagnetic field inside the TLR may be treated as several
boson modes whose frequencies are discretely distributed \cite{blais_cavity_2004,wallraff_strong_2004}.
Only one of the TLR modes, which is nearly resonant to the NAMRs,
can couple with the NAMRs effectively. For example, the voltage distribution
of the lowest even mode along the TLR is,
\begin{equation}
V(x)=\sqrt{\frac{\omega_{\mathrm{m}}}{cL}}\cos\frac{2\pi x}{L}\,(\hat{b}+\hat{b}^{\dagger}),
\end{equation}
 where $\omega_{\mathrm{m}}=2\pi/L\sqrt{lc}$ is the frequency of
this mode, $L$ is the length of the TLR, and $l,\, c$ are the inductance
and capacitance per unit length.

The voltage gets maximum at the two ends, where the NAMRs are coupled
with the TLR via a displacement dependent capacitance \cite{geller_superconducting_2005,sun_quantum_2006,wei_probing_2006,tian_parametric_2008,tian_ground_2009}.
The vibration mode of each NAMR may be also treated as a single boson.
To the lowest order, $C_{x}$ depends linearly on the movement of
the NAMR, $C_{x}\simeq C_{x}^{0}(1+\hat{x}/d_{0})$. Applying a voltage
bias $V_{g}$ to the NAMRs, we have the interaction as
\begin{equation}
H_{\mathrm{int}}=\frac{1}{2}C_{x}^{0}(1+\frac{\hat{x}}{d_{0}})\big(V(x)-V_{g}\big)^{2}.
\end{equation}
Quantizing the coordinate of the NAMR as $\hat{x}=\delta x_{0}(\hat{a}+\hat{a}^{\dagger})$,
we obtain an interaction term as $H_{\mathrm{int}}=g(\hat{a}+\hat{a}^{\dagger})(\hat{b}+\hat{b}^{\dagger})$.
For typical parameters, $C_{x}^{0}\simeq0.65\,\mathrm{fF}$, $V_{g}=4V$,
$d_{0}\simeq50\,\mathrm{nm}$, $\delta x_{0}\simeq5\,\mathrm{fm}$,
$lc\simeq4\,\mathrm{fF}$, $\omega_{\mathrm{m}}/2\pi\simeq5\,\mathrm{GHz}$,
the coupling strength is estimated as $g/2\pi\simeq6\,\mathrm{MHz}$
\cite{knobel_nanometre-scale_2003,tian_parametric_2008}. In this
regime, it is appropriate to apply J-C approximation to have $H_{\mathrm{int}}=g(\hat{a}\hat{b}^{\dagger}+\hat{a}^{\dagger}\hat{b})$.

\begin{figure}
	\begin{center}
	\resizebox{0.45\textwidth}{!}{
		\includegraphics{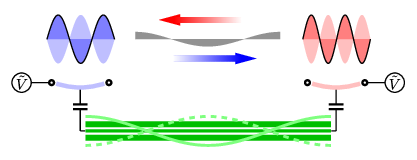}
		}
	\end{center}
\caption{(Color online) Two remotely located NAMRs are indirectly coupled via
a superconducting TLR. The bias voltage provides a difference between
the NAMR and the TLR, so they can be coupled by a capacitance. At
the same time, the voltage noise provides each NAMR with independent
heat bath which has different effective temperature.}
\end{figure}

For the mediating TLR, $\omega_{\mathrm{m}}/2\pi\simeq5\,\mathrm{GHz}$,
$Q>10^{4}$, and the lifetime of the photon inside the resonator is
$\tau>1\,\mu\mathrm{s}$ \cite{wallraff_strong_2004}. The NAMR with
$\omega/2\pi>1\,\mathrm{GHz}$ usually has a lower mechanics quality,
$Q\simeq500$ \cite{henry_huang_nanoelectromechanical_2003}, and
it depends on the fabrication techniques \cite{greenberg_nanomechanical_2012}.
Thus the relaxation time of the NAMR is much shorter than the TLR,
and we can neglect the dissipation of the TLR. The voltage noises
applied to the NAMRs at the two sides, which arise from resistance
and bring in the Joule heat, may provide the NAMRs with independent
heat baths with different effective temperatures, and this can be
controlled and measured in experiment \cite{giazotto_opportunities_2006,chen_quantum_2012,giazotto_josephson_2012}.

\section{Summary}

In summary, we have studied the long-term behavior of a coupled HO
system connecting with a complicated environment which consists of
two independent heat baths with different temperatures. We derived
a master equation with respect to the normal modes. With the help
of the characteristic description of Wigner function, we obtained
the numerical and analytical results for the steady state of each
local oscillator.

These results show that the inter-mode transitions mediated by the
environment are essential to non-equilibrium flux between the interacting
subsystems, thus they would contribute to long-term effect even in
Markovian systems. This is different from the case in conventional
thermalization problems, where only one canonical heat bath is involved.
The non-thermal stabilization process is determined by the competition
between the rate of the inter-mode transition and that of the energy
exchange with each private heat bath. 

This work is supported by National Natural Science Foundation of China under Grants Nos. 11121403, 10935010 and 11074261, National 973-program Grants No. 2012CB922104, and Postdoctoral Science Foundation of China No. 2013M530516.

\appendix

\section{Derivation of master equation}

We show the derivation of the master equation Eq.\,(\ref{eq:ME-noRWA})
here. In the interaction picture of $H_{S}+H_{B}$, the interaction
with the environment becomes,
\begin{align}
V_{I}(t) & =V_{I}^{L}(t)+V_{I}^{R}(t),\label{eq:app-VI}\\
V_{I}^{\sigma}(t) & =\hat{a}_{\sigma}^{\dagger}(t)\Gamma_{\sigma}(t)+\hat{a}_{\sigma}(t)\Gamma_{\sigma}^{\dagger}(t),\nonumber 
\end{align}
 where 
\begin{align}
\hat{a}_{\sigma}(t) & =\sum_{j}U_{j\sigma}^{*}\hat{A}_{j}e^{-i\varepsilon_{j}t},\label{eq:app-A=00003Dua}\\
\Gamma_{\sigma}(t) & =\sum_{\mathbf{k}_{\sigma}}g_{\mathbf{k}_{\sigma}}\,\hat{c}_{\mathbf{k}_{\sigma}}e^{-i\omega_{\mathbf{k}_{\sigma}}t}.\nonumber 
\end{align}
Here $\hat{A}_{i}=U_{ij}\hat{a}_{j}$ are the normal modes of the
interacting oscillators system.

We take Born-Markovian approximation and put these interaction terms
into the following equation \cite{breuer_theory_2002},

\begin{align}
\partial_{t}\rho= & -\int_{0}^{\infty}d\tau\,\mathbf{Tr}_{B}\left[V_{I}(t),\left[V_{I}(t-\tau),\rho(t)\otimes\rho_{B}\right]\right]\label{eq:redfield-1}\\
= & -\int_{0}^{\infty}d\tau\,\mathbf{Tr}_{B}\left[V_{I}^{L}(t),\left[V_{I}^{L}(t-\tau),\rho(t)\otimes\rho_{B}\right]\right]\nonumber \\
 & -\int_{0}^{\infty}d\tau\,\mathbf{Tr}_{B}\left[V_{I}^{R}(t),\left[V_{I}^{R}(t-\tau),\rho(t)\otimes\rho_{B}\right]\right].\nonumber 
\end{align}
Here we assume that the state of each bath is a canonical thermal
one and does not change with time, i.e., $\rho_{B}=\rho_{B}^{L}\otimes\rho_{B}^{R}$,
$\rho_{B}^{\sigma}\propto\exp[-H_{B}^{\sigma}/kT_{\sigma}]$ and $H_{B}^{\sigma}=\omega_{\mathbf{k}_{\sigma}}\hat{c}_{\mathbf{k}_{\sigma}}^{\dagger}\hat{c}_{\mathbf{k}_{\sigma}}$
where $T_{L/R}$ is the temperature of the left/right heat bath. Thus,
terms like $\mathbf{Tr}_{B}[V_{I}^{L}(t)V_{I}^{R}(t-\tau)\rho(t)\otimes\rho_{B}]$
always vanish, because they only contain the first moment of each
bath.

The rightside of Eq.\,(\ref{eq:redfield-1}) contains two integrals
of the same form. Each integral gives four terms, one of which is
calculated bellow as an example, 
\begin{align}
 & \int_{0}^{\infty}d\tau\,\mathbf{Tr}_{B}\Big[\hat{a}_{\sigma}^{\dagger}(t)\Gamma_{\sigma}(t)\cdot\rho(t)\otimes\rho_{B}\cdot\hat{a}_{\sigma}(t-\tau)\Gamma_{\sigma}^{\dagger}(t-\tau)\Big]\nonumber \\
= & \int_{0}^{\infty}d\tau\,\hat{a}_{\sigma}^{\dagger}(t)\rho(t)\hat{a}_{\sigma}(t-\tau)\big<\Gamma_{\sigma}^{\dagger}(t-\tau)\Gamma_{\sigma}(t)\big>_{B},\label{eq:integral}
\end{align}
where 
\begin{align}
\hat{a}_{\sigma}^{\dagger}(t)\rho(t)\hat{a}_{\sigma}(t-\tau)= & \sum_{i,j}U_{i\sigma}U_{j\sigma}^{*}\,\hat{A}_{i}^{\dagger}\rho\hat{A}_{j}\, e^{i(\varepsilon_{i}-\varepsilon_{j})t}\cdot e^{i\varepsilon_{j}\tau},\nonumber \\
\big<\Gamma_{\sigma}^{\dagger}(t-\tau)\Gamma_{\sigma}(t)\big>_{B}= & \sum_{\mathbf{k}_{\sigma}}\left|g_{\mathbf{k}_{\sigma}}\right|^{2}\langle\hat{c}_{\mathbf{k}_{\sigma}}^{\dagger}\hat{c}_{\mathbf{k}_{\sigma}}\rangle_{\mathrm{th}}\, e^{-i\omega_{\mathbf{k}_{\sigma}}\tau}\\
= & \int_{0}^{\infty}d\omega\, J_{\sigma}(\omega)N_{\sigma}(\omega)\, e^{-i\omega\tau}.\nonumber 
\end{align}
Here, $J_{\sigma}(\omega)=\sum_{\mathbf{k_{\sigma}}}\left|g_{\mathbf{k}_{\sigma}}\right|^{2}\delta(\omega-\omega_{\mathbf{k}_{\sigma}})$
is the coupling spectrum, and $N_{\sigma}(\varepsilon_{i})=[\exp(\varepsilon_{i}/kT_{\sigma})-1]^{-1}$
is the Planck distribution with temperature $T_{\sigma}$. The integral
Eq.\,(\ref{eq:integral}) gives 
\begin{multline}
\sum_{i,j}U_{i\sigma}U_{j\sigma}^{*}\,\hat{A}_{i}^{\dagger}\rho\hat{A}_{j}\, e^{i(\varepsilon_{i}-\varepsilon_{j})t}\\
\times\frac{\gamma_{\sigma}(\varepsilon_{j})}{2}\cdot N(\varepsilon_{j})+i\mathbf{P}\int_{0}^{\infty}d\omega\,\frac{J_{\sigma}(\omega)N_{\sigma}(\omega)}{\varepsilon_{j}-\omega}.
\end{multline}
Here, we denote $\gamma_{\sigma}(\varepsilon_{i})=2\pi J_{\sigma}(\varepsilon_{i})$,
which characterizes the coupling strength with each heat bath. The
principle integral contributes to Lamb shift, and we omit this term
in this paper.

The physical meaning of Eq.\,(\ref{eq:integral}) may be understood
in the following way. At time $t-\tau$, the coupled HO system emits
energy to the environment, and then absorbs back at time $t$. However,
the energy exchange with the environment during this process is done
by the total normal modes $\hat{A}_{i}$ but not the local modes $\hat{a}_{i}$.
Thus, when the emission and absorption modes are not the same one,
there is an oscillating factor $\exp[i(\varepsilon_{i}-\varepsilon_{j})t]$
left. $\delta\epsilon_{ij}\equiv\varepsilon_{i}-\varepsilon_{j}$
characterizes the splitting amplitude resulting from the coupling.
By the mediation of the environment, the different normal modes $\hat{A}_{i}$
of the system are coupled together.

Other terms of Eq.\,(\ref{eq:redfield-1}) can be also obtained as
above. Each of the two integrals gives the following Lindblad-like
form with an extra oscillating factor, 

\begin{align}
 & \sum_{ij}\frac{1}{2}U_{i\sigma}U_{j\sigma}^{*}\Big(\gamma_{\sigma}(\varepsilon_{i})[N_{\sigma}(\varepsilon_{i})+1]+\gamma_{\sigma}(\varepsilon_{j})[N_{\sigma}(\varepsilon_{j})+1]\Big)\nonumber \\
 & \qquad\times\Big(\hat{A}_{j}\rho\hat{A}_{i}^{\dagger}-\frac{1}{2}\{\hat{A}_{i}^{\dagger}\hat{A}_{j},\,\rho\}_{+}\Big)e^{i\delta\epsilon_{ij}t}\nonumber \\
 & +\sum_{ij}\frac{1}{2}U_{i\sigma}U_{j\sigma}^{*}[\gamma_{\sigma}(\varepsilon_{j})\, N_{\sigma}(\varepsilon_{i})+\gamma_{\sigma}(\varepsilon_{j})\, N_{\sigma}(\varepsilon_{j})]\label{eq:lindblad-1}\\
 & \qquad\times\Big(\hat{A}_{i}^{\dagger}\rho\hat{A}_{j}-\frac{1}{2}\{\hat{A}_{j}\hat{A}_{i}^{\dagger},\,\rho\}_{+}\Big)e^{i\delta\epsilon_{ij}t}.\nonumber 
\end{align}

For simplicity, we assume $\gamma_{\sigma}(\varepsilon_{i})\simeq\gamma_{\sigma}$
does not depend too much on $\omega$ and can be treated as constant.
In sum of Eqs.\,(\ref{eq:redfield-1}, \ref{eq:lindblad-1}), we
get the following master equation in Schr\"odinger's picture, and
the oscillating factors do not appear,
\begin{align}
\partial_{t}\rho= & i[\rho,\,\sum\varepsilon_{i}\hat{A}_{i}^{\dagger}\hat{A}_{i}]\nonumber \\
 & +\sum_{ij}\,\frac{\Lambda_{ij}^{-}}{2}\big(2\hat{A}_{i}\rho\hat{A}_{j}^{\dagger}-\{\hat{A}_{j}^{\dagger}\hat{A}_{i},\rho\}_{+}\big)\\
 & +\frac{\Lambda_{ij}^{+}}{2}\big(2\hat{A}_{i}^{\dagger}\rho\hat{A}_{j}-\{\hat{A}_{j}\hat{A}_{i}^{\dagger},\rho\}_{+}\big),\nonumber 
\end{align}
 where 
\begin{align}
\Lambda_{ij}^{+}= & \frac{\gamma_{L}}{2}U_{i1}U_{j1}^{*}[N_{L}(\varepsilon_{i})+N_{L}(\varepsilon_{j})]\nonumber \\
 & +\frac{\gamma_{R}}{2}U_{i3}U_{j3}^{*}[N_{R}(\varepsilon_{i})+N_{R}(\varepsilon_{j})],\\
\Lambda_{ji}^{-}= & \frac{\gamma_{L}}{2}U_{i1}U_{j1}^{*}[N_{L}(\varepsilon_{i})+N_{L}(\varepsilon_{j})+2]\nonumber \\
 & +\frac{\gamma_{R}}{2}U_{i3}U_{j3}^{*}[N_{R}(\varepsilon_{i})+N_{R}(\varepsilon_{j})+2].\nonumber 
\end{align}

The terms with $i\neq j$ describes the transition between different
normal modes. These terms are often omitted by RWA.

\section{Characteristic function of Wigner representation}

The Wigner representation often give us great convenience to study
properties of quantum oscillators. It can be defined from a characteristic
function \cite{gardiner_quantum_2004}, 
\begin{equation}
\chi_{w}(\kappa,\kappa^{*})=\mathbf{Tr}\big[e^{\kappa\hat{a}^{\dagger}-\kappa^{*}\hat{a}}\rho\big]
\end{equation}
The Wigner function is defined as the Fourier transform of $\chi_{w}(\kappa,\kappa^{*})$,
\begin{equation}
W(\alpha,\alpha^{*})=\frac{1}{\pi^{2}}\int d^{2}\kappa\, e^{-\kappa\alpha^{*}+\kappa^{*}\alpha}\chi_{w}(\kappa,\kappa^{*}).
\end{equation}

For a system that consists of two oscillators, the characteristic
function can be defined as, 
\[
\chi_{12}(\kappa_{1},\kappa_{2})=\mathbf{Tr}_{12}\big[e^{\kappa_{1}\hat{a}_{1}^{\dagger}-\kappa_{1}^{*}\hat{a}_{1}}\cdot e^{\kappa_{2}\hat{a}_{2}^{\dagger}-\kappa_{2}^{*}\hat{a}_{2}}\rho_{12}\big].
\]

From this definition, immediately we can find that if we had known
$\chi_{12}(\kappa_{1},\kappa_{2})$ for the whole system explicitly,
it would be quite easy to get the description of the subsystems $\chi_{1(2)}$,
just by setting $\kappa_{2(1)}=0$ in $\chi_{12}(\kappa_{1},\kappa_{2})$,
without having to calculate the reduced density matrix of subsystems
$\rho_{1(2)}$. This provides a simple method for us to study the
state of subsystems.

Besides, for the thermal state of a  oscillator $\rho_{T}={\cal Z}^{-1}\exp[-\frac{\omega}{kT}\hat{a}^{\dagger}\hat{a}]$,
the characteristic function is,
\begin{equation}
\chi_{T}(\kappa,\kappa^{*})=\exp\big[-(N+\frac{1}{2})\left|\kappa\right|^{2}\big].\label{eq:X-thermal}
\end{equation}
 Here $N=\left[\exp(\omega/kT)-1\right]^{-1}$ is the Planck distribution.

As seen from the definition, $\chi_{w}(\kappa,\kappa^{*})$ and $W(\alpha,\alpha^{*})$
can be mapped into each other through Fourier transformation. It is
also well known that there is one-to-one correspondence between a
physical Wigner function and a density matrix. Therefore, there is
one and only one density matrix $\rho$ decided by a legal $\chi_{w}(\kappa,\kappa^{*})$.

Thus, if we have a characteristic function $\chi(\kappa,\kappa^{*})$
which has a Gaussian form like Eq.(\ref{eq:X-thermal}), with $N\ge0$,
we can always come into the fact that the corresponding density matrix
is 
\begin{equation}
\rho=\frac{1}{{\cal Z}}\sum_{n=0}^{\infty}e^{-n\,\beta^{\mathrm{eff}}\omega}|n\rangle\langle n|,
\end{equation}
 where $\beta^{\mathrm{eff}}$ comes from $N=\left[\exp(\beta^{\mathrm{eff}}\omega)-1\right]^{-1}$.
This is a canonical state for the  oscillator with an effective temperature
$1/\beta^{\mathrm{eff}}$. A more rigorous proof lies bellow.

\emph{Proof:} From the definition of $\chi_{w}(\kappa,\kappa^{*})$,
we have 
\begin{align}
\chi_{w}(\kappa,\kappa^{*}) & =e^{\frac{1}{2}\left|\kappa\right|^{2}}\mathbf{Tr}\big[e^{\kappa\hat{a}^{\dagger}}\rho e^{-\kappa^{*}\hat{a}}\big]\\
 & =e^{\frac{1}{2}\left|\kappa\right|^{2}}\int\frac{d^{2}\alpha}{\pi}\, e^{\kappa\alpha^{*}-\kappa^{*}\alpha}\langle\alpha|\rho|\alpha\rangle.\nonumber 
\end{align}
 If we have a Gaussian formed characteristic function like Eq.(\ref{eq:X-thermal}),
we can correspondingly get $\langle\alpha|\rho_{T}|\alpha\rangle$
by reversed transformation of the equation above,
\begin{align}
\langle\alpha|\rho_{T}|\alpha\rangle & =\int\frac{d^{2}\kappa}{\pi}\, e^{-\kappa\alpha^{*}+\kappa^{*}\alpha}\cdot\exp\big[-(N+1)\left|\kappa\right|^{2}\big]\nonumber \\
 & =\frac{1}{N+1}\exp\big[-\frac{\left|\alpha\right|^{2}}{N+1}\big].\label{eq:Q-rep}
\end{align}
 On the other hand, we can also expand $\langle\alpha|\rho_{T}|\alpha\rangle$
as,
\begin{align}
\langle\alpha|\rho_{T}|\alpha\rangle & =\sum_{m,n}\left\langle \alpha|m\right\rangle \left\langle m\left|\rho_{T}\right|n\right\rangle \left\langle n|\alpha\right\rangle \\
 & =\sum_{m,n}\left\langle m\left|\rho_{T}\right|n\right\rangle \cdot\frac{(\alpha^{*})^{m}\alpha^{n}}{\sqrt{m!n!}}e^{-\left|\alpha\right|^{2}}.\nonumber 
\end{align}
Comparing with the expansion of Eq.\,(\ref{eq:Q-rep}), we can get
the matrix elements of $\rho_{T}$,
\begin{equation}
\left\langle m\left|\rho_{T}\right|n\right\rangle =\cfrac{(1+\frac{1}{N})^{-n}}{N+1}\delta_{mn}.
\end{equation}
 Denote $\exp[\beta^{\mathrm{eff}}\omega]=1+\frac{1}{N}$, we can
see that $\rho_{T}$ is a canonical state. \hfill$\blacksquare$

\section{Steady solution of Fokker-Planck equation}

The standard form of Fokker-Planck equation and its characteristic
equation are as follows,
\begin{align}
\frac{\partial P}{\partial t}+\sum_{i}\lambda_{i}\frac{\partial}{\partial y_{i}}(y_{i}P) & =\frac{1}{2}\sum_{ij}\sigma_{ij}\frac{\partial^{2}P}{\partial y_{i}\partial y_{j}},\\
\frac{\partial f}{\partial t}-\sum_{i}\lambda_{i}\xi_{i}\frac{\partial f}{\partial\xi_{i}} & =-\frac{1}{2}f\sum_{ij}\sigma_{ij}\xi_{i}\xi_{j},\quad\mathbf{Re}\lambda_{i}<0.\nonumber 
\end{align}
$f(\vec{\xi},t)$ is the Fourier transformation of $P(\vec{y},t)$,
\[
f(\vec{\xi},t)=\int d^{n}y\, P(\vec{y},t)e^{-i\vec{\xi}\cdot\vec{y}}.
\]
The equation of $f(\vec{\xi},t)$ is a first-order quasi-linear partial
differential one. It can be solved analytically \cite{wang_theory_1945},
and the solution is,
\begin{equation}
f(\vec{\xi},t)=\Phi(\xi_{i}e^{\lambda_{i}t},\cdots)\cdot\exp\Big[-\frac{1}{2}\sum_{ij}\sigma_{ij}\frac{\xi_{i}\xi_{j}}{\lambda_{i}+\lambda_{j}}\Big].
\end{equation}
$\Phi(\cdots)$ is determined according to the initial condition,
and $\Phi(t\rightarrow\infty)=1$.

In our problem, the equation of the characteristic function is
\begin{equation}
\partial_{t}\chi+\vec{z}\cdot\mathbf{T}\cdot\frac{\partial}{\partial\vec{z}^{T}}\chi=\vec{z}\cdot\mathbf{D}\cdot\vec{z}^{T}\chi.
\end{equation}
 The only difference with the standard form is that $\mathbf{T}$
is not diagonal here. We first diagonalize it and make it a standard
form. Denoting $V\cdot\mathbf{T}\cdot V^{-1}=\mathrm{diag}\{\mathbf{d}_{1},\cdots,\mathbf{d}_{n}\}$
and $\vec{z}=\vec{\xi}\cdot V$, we can transform our equation into
the standard Fokker-Planck form,
\begin{equation}
\partial_{t}\chi+\vec{\xi}\cdot\mathbf{d}\cdot\frac{\partial}{\partial\vec{\xi}^{T}}\chi=\vec{\xi}\cdot V\mathbf{D}V^{T}\cdot\vec{\xi}^{T}\chi.
\end{equation}
Now we could write down the steady solution as 
\begin{gather}
\chi(\vec{z})=\exp\left[\vec{z}V^{-1}\cdot\mathbf{D}'\cdot(\vec{z}V^{-1})^{T}\right],\\
\text{where }\qquad\mathbf{D}'_{ij}=\frac{[V\mathbf{D}V^{T}]_{ij}}{\mathbf{d}_{i}+\mathbf{d}_{j}}.\nonumber 
\end{gather}

\end{document}